\documentclass[prl,superscriptaddress, preprintnumbers,twocolumn,showpacs]{revtex4}
\usepackage{amsmath}
\usepackage{amssymb}
\usepackage{amsthm}
\usepackage{amsfonts}
\usepackage{algorithmic}
\usepackage{enumerate}
\usepackage{latexsym}
\usepackage{graphicx}
\usepackage{bm}
\usepackage{subfigure}
\usepackage{dcolumn}

\begin{document}


\title{Pure spin current in a two-dimensional topological insulator}

\author{Xing-Tao An}
\email{anxingtao@semi.ac.cn}
\affiliation{SKLSM, Institute of Semiconductors, Chinese Academy of Sciences, P. O. Box 912, Beijing 100083, China}
\affiliation{School of Sciences, Hebei University of Science and Technology, Shijiazhuang, Hebei 050018,
China}
\author{Yan-Yang Zhang}
\affiliation{SKLSM, Institute of Semiconductors, Chinese Academy of Sciences, P. O. Box 912, Beijing 100083, China}
\author{Jian-Jun Liu}
\affiliation{Physics Department, Shijiazhuang University, Shijiazhuang
050035, China}
\author{Shu-Shen Li}
\affiliation{SKLSM, Institute of Semiconductors, Chinese Academy of Sciences, P. O. Box 912, Beijing 100083, China}

\date{\today}

\begin{abstract}
We predict a mechanism to generate a pure spin current in a two-dimensional topological insulator. As the magnetic impurities exist on one of edges of the two-dimensional topological insulator, a gap is opened in the corresponding gapless edge states but another pair of gapless edge states with opposite spin are still protected by the time-reversal symmetry. So the conductance plateaus with the half-integer values $e^2/h$ can be obtained in the gap induced by magnetic impurities, which means that the pure spin current can be induced in the sample. We also find that the pure spin current is insensitive to weak disorder. The mechanism to generate pure spin currents is generalized for two-dimensional topological insulators.
\end{abstract}

\pacs{75.76.+j; 72.80.Vp; 03.65.Vf}
\keywords{Spin current; Topological insulator}
\maketitle

Since graphene, a single-layer honeycomb lattice of carbon atoms,
has been prepared laboratorially by Novoselov et al.,\cite{1} it
has attracted considerable attentions due to its novel properties in
condensed matter physics and potential applications in devices.
\cite{2,3,4,5,6,7,8,9,10} Graphene is the first independent
two-dimensional (2D) crystal that has been experimentally achieved,
which leads to new interests in 2D systems. For example, the
Kane-Mele model, a quantum spin Hall effect (QSHE) was first
proposed in graphene with spin-orbital coupling, which is the first
example of topological insulator.\cite{14,18} Topological
insulators are time-reversal symmetric systems whose intrinsic
spin-orbit coupling (SOC) opens a bulk gap while generating the
Kramers doublet of edge states owing to the nontrivial $Z_2$
invariants of the occupied bands. The edge states force electrons
with opposite spin to flow in opposite directions along the edges of
the sample, which lead to quantized spin Hall conductance. However,
the intrinsic SOC in realistic graphene is quite weak and the gap
opening was small, so the QSHE in graphene is difficult to be
observed.\cite{19} Nevertheless, recently a monolayer honeycomb
lattice of silicon called silicene has been synthesized and attracts
much attention.\cite{20, 21, 22, 23} Silicene has a relatively
large intrinsic spin-orbit gap of $1.55meV$, as makes experimentally
accessible the Kane-Mele type QSHE.\cite{22}

The development of the topological insulator opens a new and
powerful way for the spintronic applications due to its
spin-dependent edge states. How to generate pure spin currents in
low-dimensional systems is the main challenge in the field of
spintronics. The aim of this work is to propose a method for
generating a pure spin current in a 2DTI. In this Letter, as a
concrete example, we theoretically study the electron transport in
Kane-Mele model with magnetic doping at one edge, as shown in Fig.
1. Most of the results are also applicable to general 2DTIs. Before
presenting our detailed calculations, we first analyze why pure spin
current can be generated in the present device. The intrinsic SOC
which originates from intra-atomic SOC, converts the sample into a
topological insulator with a QSHE.\cite{14} The gapless edge states
are protected by time-reversal symmetry and is thus robust to
non-magnetic impurities that do not break this symmetry. But a pair
of edge states are destroyed when magnetic impurities exist on this
corresponding edge (see Fig. 1). Because another pair of edge
states, with opposite spins containing opposite propagation
directions, are still protected by time-reversal symmetry, we can
observe a pure spin current in the sample, which is confirmed by our
following calculations.

In the tight-binding representation, we consider the Kane-Mele
Hamiltonian defined on a honeycomb lattice:\cite{14,18}
\begin{eqnarray}
H&=&t\sum_{\langle{ij}\rangle,\sigma}c_{i\sigma}^{\dag}c_{j\sigma}+i\lambda\sum_{\langle\langle{ij}\rangle\rangle,\sigma}\nu_{ij}c_{i\sigma}^{\dag}s_{z}c_{j\sigma}\nonumber\\
&+&i\alpha\sum_{\langle{ij}\rangle,\sigma}c_{i\sigma}^{\dag}(\textbf{s}\times\textbf{d}_{ij})_{z}c_{j\sigma}+\lambda_{\nu}\sum_{i,\sigma}\xi_{i}c_{i\sigma}^{\dag}c_{i\sigma}\label{eq1}.
\end{eqnarray}
The symbols $\langle{ij}\rangle$ and
$\langle\langle{ij}\rangle\rangle$ denote the nearest and the
next-nearest neighbors, respectively, and
$\sigma=\uparrow,\downarrow$ (or $\pm1$) denotes spin index. The
first term is the nearest-neighbor hopping. The second term
describes the intrinsic SOC. Here the site-dependent Haldane phase
factor\cite{14} $\nu_{ij}$ is defined as
$\nu_{ij}=(\textbf{d}_{1}\times\textbf{d}_{2})/|\textbf{d}_{1}\times\textbf{d}_{2}|=\pm{1}$,
where $\textbf{d}_{i}$ denotes the vector from one atom to one of
its nearest neighbors. $s_{z}$ is a Pauli matrix describing the
electron's spin. The third term is a nearest neighbor Rashba SOC
term, which can be produced by applying an electric field
perpendicular to the sheet. The fourth term is a staggered
sublattice potential $(\xi_{i}=\pm{1})$. It is interesting to notice
that, Eq. (1) is almost applicable to silicene except for the Rashba
SOC term, which is present between the next-nearest neighbors in
silicene but between the nearest neighbors in our model. The focus
of this work is the introduction of magnetic impurities (red square
dots in Fig. 1) on the uppermost zigzag chain of the sample $(n=1)$,
\begin{equation}
\sum_{i\in\{n=1\},\sigma}\sigma{M}c_{i\sigma}^{\dag}c_{i\sigma}
\end{equation}
where $M$ is the strength of exchange interaction induced by the
magnetic impurities. This term breaks the local time reversal
symmetry on the upper edge.
\begin{figure}[htb]
\centering
\includegraphics[scale=0.8,angle=0]{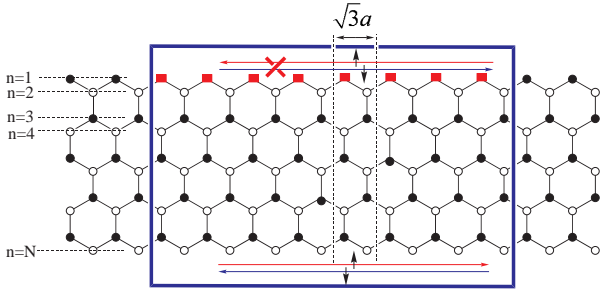}
\caption{(Color online) Schematic diagram of a zigzag honeycomb lattice nanoribbon with magnetic impurities (red square dots) on the upper edge. Its unit cell is marked by the dashed lines. The red and blue arrows at the edges denote propagation directions of the opposite spins in the edge states.}
\label{figone}
\end{figure}

In the following numerical calculations, we use the hopping energy
$t$ as the energy unit. The width $N$ is chosen as $N=50$ in all
calculations and the nearest neighbor atom-atom distance is $a$. The
strengths of the intrinsic SOC, the Rashba SOC and the staggered
sublattice potential are $\lambda=0.06t$, $\alpha=0.05t$, and
$\lambda_{\nu}=0.1t$, respectively. These parameters define the
system as a two-dimensional topological insulator.\cite{14}
\begin{figure}[htb]
\includegraphics[scale=0.6,angle=0]{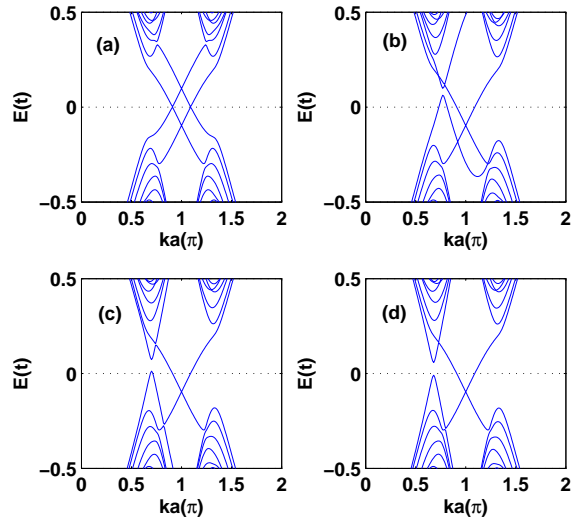}
\centering \caption{(Color online) Calculated energy bands in the
zigzag honeycomb lattice nanoribbon in the presence of the intrinsic
spin-orbit coupling and the Rashba spin-orbit coupling for $M=0$
(a), $M=0.4t$ (b), $M=0.8t$ (c), and $M=1.0t$ (d).} \label{figtwo}
\end{figure}

First, we investigate the energy subbands obtained by solving the
lattice model in strip geometry, as shown in Fig. 2. In the pure
case ($M=0$, Fig. 2(a)), the edge states traverse the energy gap in
pairs. The gapless edge states are robust against small non-magnetic
perturbations since they are protected by time reversal symmetry.\cite{14}
However, in the presence of magnetic impurities
($M\neq0$, Fig. 2(b)-(d)) at the upper edge, the corresponding pair
of gapless edge states is destroyed and a gap can be opened due to
the local time reversal symmetry breaking. Moreover, the magnitude
of the gap opened by the magnetic impurities increases with enhanced
$M$, and can reach $0.18eV$ for $M=1.0t$. Another pair of gapless
edge states, still protected by time reversal symmetry, persist on
that edge without magnetic impurities. But the electron-hole
symmetry is broken in the preserved gapless edge states, which cross
at $ka=\pi$ even for a very small value of $M$.
\begin{figure}[htb]
\centering
\includegraphics[scale=0.4,angle=0]{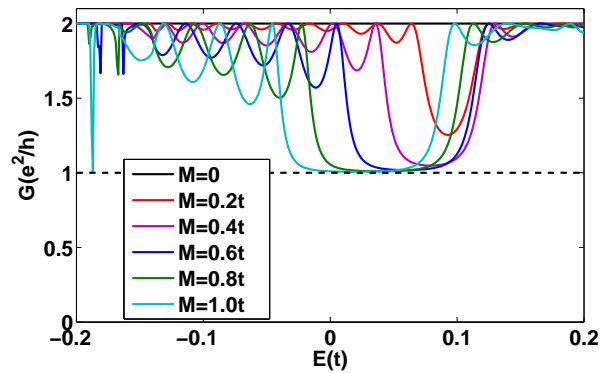}
\caption{(color online) The conductance $G$ vs $E$ for different $M$.}
\label{figthree}
\end{figure}

Next, we examine the influence of the magnetic impurities on the
conductance of the system, as shown in Fig. 3. The two-terminal
conductance of the system can be calculated by the nonequilibrium
Green's function method and the Landauer-B\"{u}ttiker formula
$G(E)=\frac{e^{2}}{h}\mathrm{Tr}[\mathbf{\Gamma}_{L}(E)\textbf{G}^{r}(E)\mathbf{\Gamma}_{R}(E)\textbf{G}^{a}(E)]$,
where
$\mathbf{\Gamma}_{p}(E)=i[\mathbf{\Sigma}_{p}^{r}(E)-\mathbf{\Sigma}_{p}^{a}(E)]$
is the line-width function and
$\textbf{G}^{r}(E)=[\textbf{G}^{a}(E)]^{\dag}=1/[\mathbf{E}-\textbf{H}_{cen}-\mathbf{\Sigma}_{L}^{r}-\mathbf{\Sigma}_{R}^{r}]$
is the retarded Green function with the Hamiltonian in the center
region $\textbf{H}_{cen}$.\cite{24} The self-energy
$\mathbf{\Sigma}_{p}^{r}$ due to the semi-infinite lead-$p$ can be
calculated numerically.\cite{25} In the case of $M=0$, the
quantized conductance plateau appears, with the plateau value
$2e^{2}/h$ coming from the contributions of two pairs of the gapless
edge states. For greater values of $M$, the conductance plateau
$2e^2/h$ is suppressed and evolves into a conductance plateau
$e^2/h$ in the energy gap opened by the magnetic impurities. The
conductance plateau $e^2/h$ is only contributed from the gapless
edge states at the lower edge without magnetic impurities, which can
be seen from Fig. 4. Without magnetic impurities ($M=0$), there are
two pairs of perfect edge states with opposite spins at two edges of
the sample, so the currents through the sample is spin unpolarized.
However, when the gapless edge states at the upper edge are
destroyed by magnetic impurities ($M\neq0$), the spin-velocity
locked channels persist only at the lower edge of the sample.
Therefore, the current in the sample consists of the opposite spins
moving in opposite directions that is called pure spin current,
which is just the aim of the present work. For a definite
arrangement of bias voltage, there remains only, say, right going
channels with spin up working. Moreover, the pure spin current in
the sample is really invariant under time reversal.

\begin{figure}[htb]
\centering
\includegraphics[scale=0.8,angle=0]{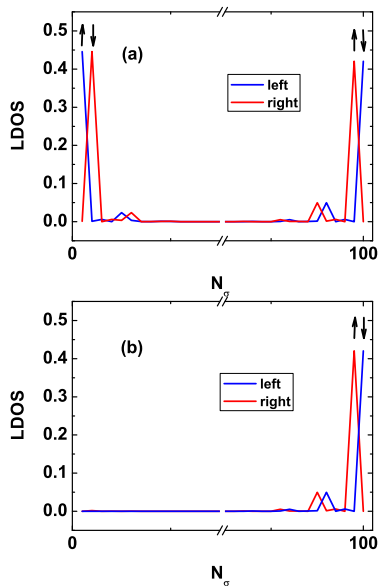}
\caption{(Color online) Spin-resolved local density of states (LDOS) along the width of sample for (a) $M=0$ and (b) $M=1.0$. The energy is chosen $E=0.02t$ in the simulation. $N_{\sigma}=2n-1$ and $N_{\sigma}=2n$ correspond to spin-up and spin-down LDOS, respectively, where $N_{\sigma}=1, 2, \cdots, 2N$. The black up and down arrows denote spin up and spin down. The blue and red curves denote left and right propagation direction}
\label{figfour}
\end{figure}
To test the above arguments in a more direct way, we also studied
the spin-resolved conductance and spin polarization when the
magnetic impurities exist on the upper edge of the sample. For the
sake of simplicity, we assume that in the leads, the staggered
sublattice potential, the intrinsic and Rashba SOC do not exist,
i.e., the Hamiltonian of lead- $p$ is simply
\begin{eqnarray}
H_{p}=t\sum_{\langle{ij}\rangle,\sigma}c_{i\sigma}^{\dag}c_{j\sigma}.
\end{eqnarray}
The spin-resolved conductance matrix can be written as
\begin{eqnarray}
G=\left(
\begin{array}{ccc}
    G_{\uparrow\uparrow} & G_{\uparrow\downarrow}\\
    G_{\downarrow\uparrow} & G_{\downarrow\downarrow}\\
  \end{array}
\right),
\end{eqnarray}
which can also be calculated by generalized Landauer formula for
spin transport. The conductance $G_{\uparrow\uparrow}$ and
$G_{\uparrow\downarrow}$ can be obtained when we assume that only
spin-up electrons are injected from the left lead into the sample
and collected in the right lead. We can also calculate
$G_{\downarrow\uparrow}$ and $G_{\downarrow\downarrow}$ in the same
way by assuming only spin-down electrons are injected from the left
lead. The total conductance $G$ and the spin polarization $P$ in
lead-$R$ can be respectively defined as\cite{26,28}
\begin{eqnarray}
G=G_{\uparrow\uparrow}+G_{\downarrow\uparrow}+G_{\uparrow\downarrow}+G_{\downarrow\downarrow}
\end{eqnarray}
and
\begin{eqnarray}
P=\frac{G_{\uparrow\uparrow}+G_{\downarrow\uparrow}-G_{\uparrow\downarrow}-G_{\downarrow\downarrow}}{G_{\uparrow\uparrow}+G_{\downarrow\uparrow}+G_{\uparrow\downarrow}+G_{\downarrow\downarrow}}.
\end{eqnarray}

\begin{figure}[htb]
\centering
\includegraphics[scale=0.9,angle=0]{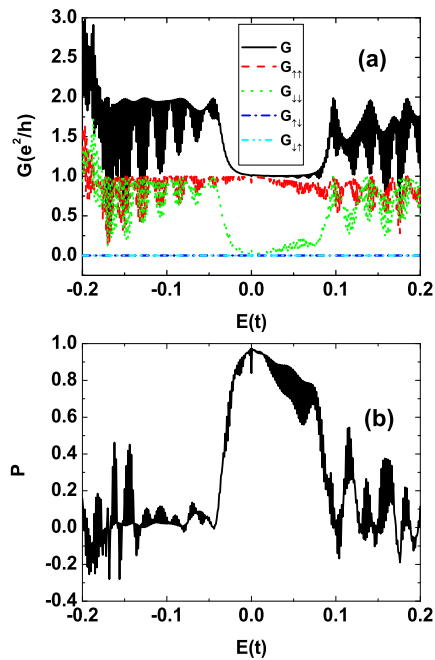}
\caption{(Color online) Spin-resolved conductance (a) and spin polarization (b) vs $E$ for $M=1.0$. The parameters $\lambda=0$, $\alpha=0$, and $\lambda_{\nu}=0$ are chosen in leads}
\label{figfour}
\end{figure}
Fig. 5 (a) and (b) show the spin-resolved conductance $G$ and spin
polarization $P$ versus the energy $E$ for $M=1$. In Fig. 5 (a), the
total conductance manifests itself with the plateau value $e^{2}/h$
due to the presence of the magnetic impurities on the upper edge of
the central region. Due to the topological nature of the edge
states, this plateau is insensitive to the mismatch between the
sample and the leads. We can also find that the spin-up and
spin-down electrons are not mixed when they transport through the
sample, i.e., $G_{\uparrow\downarrow}=G_{\downarrow\uparrow}=0$. The
spin polarization can almost reach $100\%$ in the energy gap opened
by the magnetic impurities [see Fig. 5 (b)] because the spin-up
electron can fully transport through the sample while the spin-down
electron can hardly transport through the sample in the gap [see
Fig. 5 (a)]. Beyond the gap, due to the conduction band mismatch
between the central region and the leads, there are resonant
tunneling peaks in the conductance beyond the gap induced by the
magnetic impurities, as shown in Fig. 3 and Fig. 5.
\begin{figure}[htb]
\centering
\includegraphics[scale=0.9,angle=0]{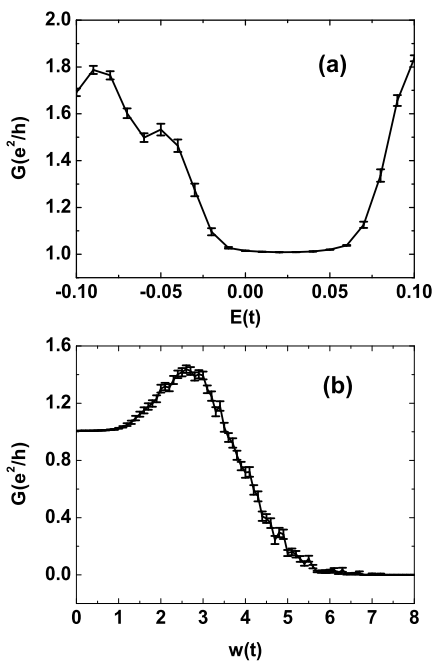}
\caption{The conductance $G$ as a function of the energy $E$ for the disorder strengths $w=0.5t$ (a) and as a function of the disorder strengths $w$ for the energy $E=0.02t$ (b). The error bars show standard deviation of the conductance for $100$ samples.}
\label{figfour}
\end{figure}

Finally, we examine the non-magnetic disorder effect on this
spin-polarized conductance plateau $e^2/h$. Due to disorder, random
on-site potential $w_{i}$ is added for each site $i$ in the central
region, where $w_{i}$ is uniformly distributed in the range $[-w/2,
w/2]$ with the disorder strength $w$. Fig. 6 (a) and (b) show the
conductance $G$ versus the energy $E$ at the disorder strength
$w=0.5t$ and $G$ versus the disorder strength $w$ at the energy
$E=0.02t$, respectively. The results show that the quantum plateau
of $e^2/h$ is very robust against non-magnetic disorder because of
the topological origin of the edge states. The quantum plateau
maintains its quantized value very well even when $w$ reaches
$1.0t$. The robust and stable plateau of $e^2/h$ means that the pure
spin current of the system is insensitive to weak disorder and
protected by time-reversal symmetry. In addition, even for a large
disorder strength $w$ (e.g., from $w=1.0t$ to $w=3.0t$), the
conductance is increased rather than decreasing with the increasing
disorder strength. This is because although the strong disorder
weakens the edge states, it also result in the mobility of the
energy band structure,\cite{27} so the value of $G$ increases in
the range of $w=1.0t$ to $w=3.0t$. With further increasing of the
disorder strength, the conductance gradually reduce to zero, the
system eventually enters the insulating regime.

In summary, we predict a new mechanism to generate a pure spin
current in a two-dimensional topological insulator. As the magnetic
impurities exist on one edge of the sample, the corresponding
gapless edge states is destroyed but another pair of gapless edge
states with opposite spin are protected by time-reversal symmetry.
So a pure spin current with the spin-up and spin-down carriers
moving in opposite directions can be observed in the system.
Moreover, the pure spin current has also been found to be robust
against non-magnetic disorder. The mechanism to generate pure spin
currents can be generalized for two-dimensional topological
insulators, such as HgTe/CdTe quantum well and silicene nanoribbons.

This work was supported by National Natural Science Foundation of
China (Grant Nos. 11047184, 11104059, No. 61176089) and Hebei
province Natural Science Foundation of China (Grant No.
A2011208010).

\end{document}